\documentclass[12pt,a4paper,final]{iopart}

\usepackage{graphicx}
\pdfoutput=1
\usepackage{amsfonts,amsmath,amsthm,amssymb,amsopn}
\usepackage{dsfont,bbm}
\usepackage{mathrsfs}
\usepackage{leftidx}
\usepackage{physics}
\usepackage{cite}
\usepackage{booktabs}
\usepackage[retainorgcmds]{IEEEtrantools}
\usepackage{soul,xcolor}
\usepackage{color}
\definecolor{darkblue}{rgb}{0.,0.,0.5}
\definecolor{darkred}{rgb}{0.6,0.,0.}
\definecolor{darkgreen}{rgb}{0.,0.8,0.}
\definecolor{darkpink}{rgb}{1,0.08,0.5}

\usepackage[breaklinks=true,colorlinks=true,linkcolor=darkblue,urlcolor=magenta,citecolor=magenta]{hyperref}
\numberwithin{equation}{section}
\usepackage{cleveref}
\def\be{\begin{equation}}
\def\ee{\end{equation}}
\def\bra#1{\mathinner{\langle{#1}|}}
\def\ket#1{\mathinner{|{#1}\rangle}}

\def\la{{\leftarrow}}

\global\long\def\al{\alpha}
\global\long\def\no{\nonumber}
\global\long\def\be{\beta}
\global\long\def\ga{\gamma}
\global\long\def\de{\delta}

\global\long\def\la{\lambda}
\global\long\def\ka{\kappa}

\global\long\def\eps{\epsilon}

\newcommand{\nc}{\newcommand}
\nc{\ir}{\mathrm{i}}
\nc{\eE}{\mathsf{e}}

\global\long\def\erw#1{\left\langle #1\right\rangle }

\eqnobysec

\begin{document}
\title{Universality in 
driven systems with a multiply-degenerate umbilic point}

\author{Johannes Schmidt}
\address{Bonacci GmbH, Robert-Koch-Str. 8, 50937 Cologne, Germany}
\author{\v Ziga Krajnik}
\address{Department of Physics, New York University, 726 Broadway, New York, NY 10003, United States}

\author{Vladislav Popkov} \address{Faculty of Mathematics and
  Physics, University of Ljubljana, Jadranska 19, SI-1000 Ljubljana,
  Slovenia} \address{Department of Physics, University of
  Wuppertal, Gaussstra\ss e 20, 42119 Wuppertal, Germany}

\begin{abstract}
We investigate a driven particle system,  a multilane asymmetric exclusion process,  where the particle number in every  lane  is  conserved,  and stationary state is fully uncorrelated.
 The phase space has,  starting from three lanes and more,   an umbilic manifold where characteristic velocities of all the modes but  one coincide,  thus allowing us to study a weakly hyperbolic  system with arbitrarily large degeneracy.  
We then study space-time fluctuations in the steady state,  at the umbilic manifold, which are expected to exhibit 
universal scaling features. We formulate an effective mode-coupling theory (MCT) for the multilane model within the umbilic subspace and test its predictions.   
Unlike in the bidirectional two-lane model with an umbilic point studied earlier,  here we find  a robust $z=3/2$ dynamical exponent for the umbilic mode.  The umbilic scaling function, obtained from Monte-Carlo simulations  appears to have a universal shape for a range of interaction parameters and depends only on umbilic mode degeneracy. Remarkably, the shape and dynamic exponent of the  non-degenerate mode can be analytically predicted  on the base of effective MCT, up to a non-universal scaling factor.
Our findings suggest the existence of novel universality classes with dynamical exponent $3/2$, appearing in long-lived hydrodynamic modes with equal characteristic velocities.
\end{abstract}


\maketitle

\pagestyle{empty}
\tableofcontents
\pagestyle{headings}

\flushbottom
\clearpage

\section{Introduction}

Systems of driven interacting diffusing particles are paradigmatic models of far from equilibrium dynamics with a broad range of applications in biological, social and physical contexts \cite{SchuetzBook,MukamelReview,SchadschneiderBook}.
A  coarse-grained (hydrodynamic) description of such models, while smearing out microscopic details, retains the most important 
properties of the microscopic dynamics such as symmetries and conservation laws.  
In particular,  local conservation laws,  under typical assumptions such as locality of interactions,stability of a homogeneous steady state,  etc,   lead to long-lived slow modes,  which  show universal 
(i.e.~dependent only on a few crucial properties, such as nonlinearity and locality) behavior at  large space-time scales.  
For instance, one-dimensional driven stochastic systems of  particles interacting via short range interactions with a single global
conservation law can be mapped to surface growth processes \cite{1997ReviewKrug} and  belong to the renowned KPZ universality class \cite{Prae04}. 
In case of several conservation laws, the temporal evolution of long-lived modes is described via nonlinear fluctuating hydrodynamics
and mode-coupling theory \cite{Spoh14}.  

A cornerstone assumption of nonlinear fluctuating hydrodynamics and mode coupling theory is a spatial separation of modes with time,  an assumption referred to as strong hyperbolicity in the theory of hyperbolic conservation laws and hyperbolic shocks \cite{Lax2006}. Strong hyperbolicity is crucial for an analysis of hyperbolic shocks, which are, in turn, crucial for understanding intrinsic features of non-equilibrium systems, with no equilibrium counterpart,  e.g.~boundary-driven phase transitions \cite{SchuetzBook}. 

At a point in phase space where characteristic velocities of two (or more) modes  coincide long-lived modes do not separate in space at long times. Such a point is refereed to as an umbilic point (UP), or a point of weak hyperbolicity. Breakdown of the strong hyperbolicity assumption at the UP  leads to drastic consequences for temporal dynamics and changes the 
types of admissible shocks, see e.g.~\cite{1995Chen,2001Chen}.  

Universal space-time  fluctuations in hyperbolic systems at large scales are also sensitive to the issue of mode separation. 
While universality properties of space-time  fluctuations in strongly hyperbolic systems are relatively well understood,  in terms of mode-coupling theory \cite{Spoh14}, only a few recent studies \cite{2024Spohn, 2025Spohn, 2025SpohnSingle,  2025UmbilicBidirectional}  treat universality aspects of the weakly hyperbolic scenario. All these recent studies   concern a scenario where  a model (either a system of stochastic differential equations, or a driven diffusive particle system) has a single  
doubly-degenerate umbilic mode, and no other conserved modes exist.  While this scenario is undoubtedly basic and minimal,  it leaves open some fundamental questions,  for instance: (i) How does a degenerate umbilic mode interact with conventional modes? (ii) Can this interaction be described by mode-coupling theory? (iii) What happens in case of a multiple (higher than double) umbilic degeneracy? 

In the present communication we, at least partly, clarify the above questions,  by considering an interacting particle model with several conservation laws, a multilane totally asymmetric exclusion process. The process is extremely simple and has a fully uncorrelated steady state,  rendering an analytic treatment straightforward and numerical simulations  efficient and easy to control.  Despite its  simplicity,  the model supports the existence of umbilic modes of arbitrary degeneracy, and additionally has a single nondegenerate mode, which separates from the umbilic mode with time. 
In the following, we recall the model's formulation, and investigate stationary density fluctuations of the umbilic mode and of the conventional nondegenerate mode both analytically (via a mode-coupling theory MCT) and numerically (via Monte-Carlo simulations).  For the conventional mode,  we find an exact correspondence with MCT predictions.  For the umbilic mode, our scaling analysis confirms a robust $z=3/2$ dynamical exponent,
expected from MCT,  and a novel family of  scaling functions,  which depend both on the interaction parameter and on the degree of degeneracy of the umbilic mode. 

\begin{figure}[ptb]
\center
\includegraphics[width=0.9\linewidth]{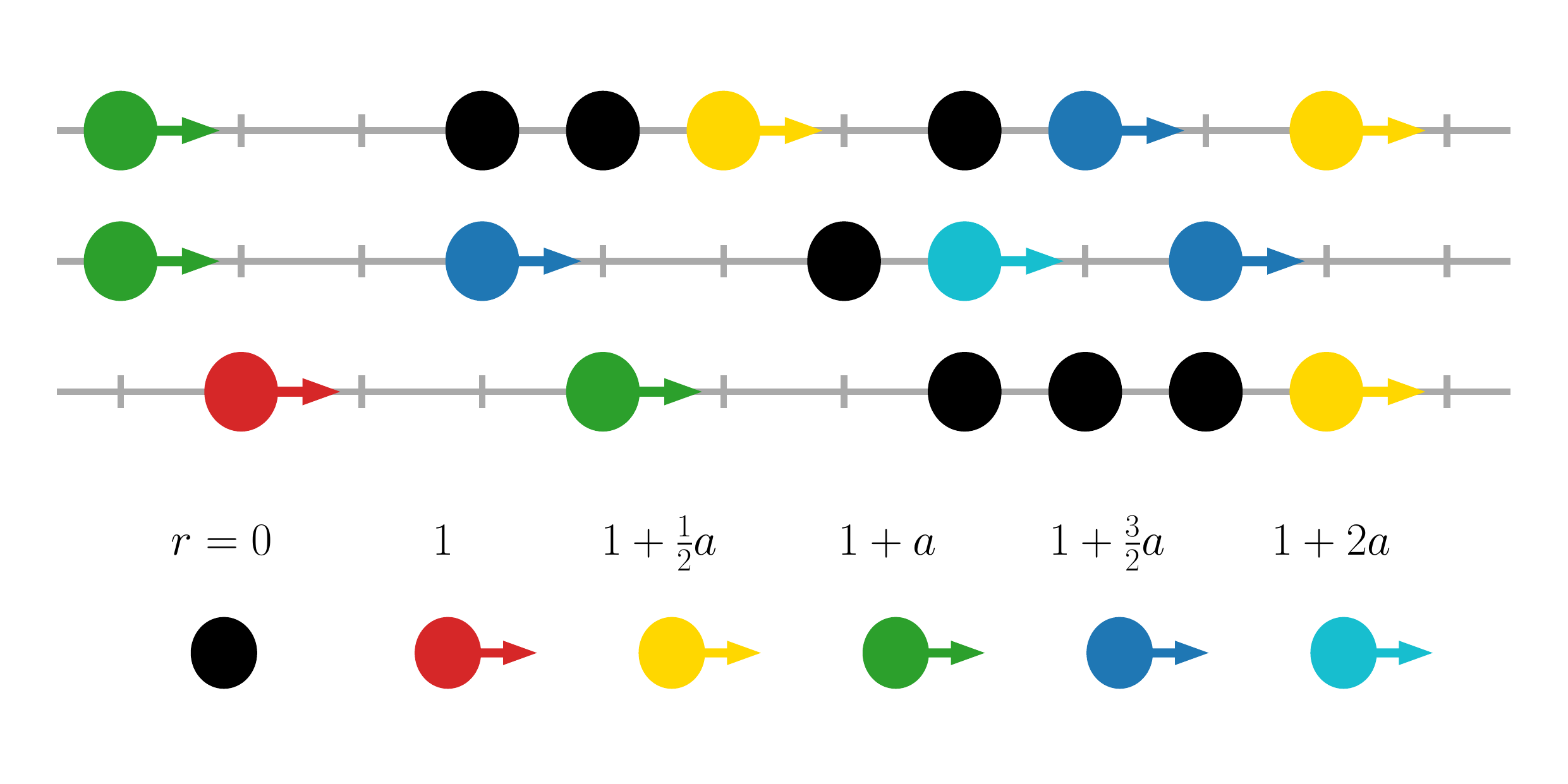}
\caption{
Allowable hops for the three-lane model ($K+1=3$) with their rates (\ref{eq:rates}).  Particles with the same color hop with the same rates,  e.g.~hopping of black-coloured particles is forbidden by hardcore exclusion. Note that all particles are identical,  and  they do not change  lanes, so that the overall particle number in each lane is conserved.
}
\label{FigMultichainTASEP}
\end{figure}

\section{Multilane TASEP: an ideal candidate for studying umbilic modes}
To observe an umbilic mode, two or more long-lived hydrodynamic modes with coinciding characteristic velocities (respectively, two or more conservation laws) are required.  For our study we use a  generalization of a totally asymmetric exclusion process (TASEP) to the multi-chain setting, proposed in \cite{2004MultilanePopkovSalerno}. The process is illustrated in Fig.~\ref{FigMultichainTASEP}.  In each lane, a particle jumps unidirectionally to the next neighboring site on the same lane with hardcore exclusion; there are $K+1$ parallel lanes; the rate of hopping depends on local occupation on the other $K$ lanes as 
\begin{equation}
r_{k}^{\lambda}=1+\frac{a}{2}\sum_{\underset{\mu\not=\lambda}{\mu=1}}^{K+1}
\left(n_{k}^{\mu}+n_{k+1}^{\mu}\right),  \label{eq:rates}
\end{equation} 
 where $n_k^\mu$ is a occupation number on site $k$ and lane number $\mu$ while $a$ is the interchain interaction. Due to hardcore exclusion sites can be either empty or occupied by one particle, $n_k^\mu=0,1$. In (\ref{eq:rates}) we have chosen a ``meanfield" reduction of the multilane TASEP model \cite{2004MultilanePopkovSalerno} where a particle in one lane is influenced by neighboring particles in all the other lanes.
A continuous time Markov process with 
rates (\ref{eq:rates})  on an infinite lattice has a spatially homogeneous, fully uncorrelated (product measure) steady state,  parametrized by set of the average particle densities in each lane $\{ \rho_\mu\}_{\mu=1}^{K+1}$.     
The static space correlation matrix in the steady state is 
\begin{equation}
K_{\la \mu} = \erw{n_{k}^{\lambda}n_{0}^{\mu}}-\rho_\la \rho_\mu = \de_{\la \mu} \de_{k,0} \,\rho_\la (1-\rho_\la).
\label{eq:Covariance}
\end{equation}
In the   hydrodynamic limit,   $n_k^\mu(t) \rightarrow \rho_\mu(x,t)$ the system dynamics on the Euler scale is given by a  system of continuity equations
\begin{align}
&\frac{\partial \rho_\la}{ \partial t}+ \frac{\partial j_\la(\rho_1,\rho_2, \ldots  ,\rho_{K+1})}{ \partial x}=0
\label{eq:continuity_eq}
\end{align}
where $j_\la(\rho_0,\ldots ) =\langle r_{k}^{\lambda}\,  n_{k}^\la (1-n_{k+1}^\la)\rangle $  is the  expectation of a particle current on lane $\la$ in the steady state,
 \begin{align}
&j_\la(\rho_1,\ldots ,\rho_{K+1}) = \rho_\la (1-\rho_\la) \left( 1 + a \sum_{\underset{\mu\not=\lambda}{\mu=1}}^{K+1}    \rho_\mu\right). \label{eq:current-densityRelation}
\end{align}
Nonnegativity of the rates (\ref{eq:rates})
implies $-1/K \leq a < \infty$.  In absence of the interchain interaction
$a=0$, the process Fig.~\ref{FigMultichainTASEP} splits into 
$K+1$ independent totally asymmetric exclusion processes (TASEPs) characterized by the KPZ 
universality \cite{KPZ,Prae04}. 
For generic  interlane interaction $a\neq 0$ we have $(K+1)$ conservation laws (\ref{eq:continuity_eq}).   The characteristic velocities, i.e.  velocities of  local perturbations over a stationary 
state with densities $\rho_{\mu}$ are given by the eigenvalues of the Jacobian $J_{\la\mu} =\frac{\partial j_\la}{ \partial \rho_\mu}$, 
\begin{align}
&J \ket{\la}=c_\la(\rho_1,\ldots, \rho_{K+1} )  \ket{\la}. \label{eq:Jeiv}
\end{align}
Interestingly, for an appropriate choice of densities, 
the multilane model of Fig.~\ref{FigMultichainTASEP} with a slight generalization of the rates (\ref{eq:rates}) 
generates a entire discrete set of Fibonacci universality classes \cite{2015Fibonacci}, with dynamical exponents given
by the ratios of nearest Fibonacci numbers $z_\al=3/2,5/3,8/5,\ldots $ \cite{2015Fibonacci}, including the Golden ratio critical exponent.  All these cases assume space separation of the long-lived modes with time,  meaning that  characteristic velocities
of all modes are distinct.

Presently, on the contrary,  we are looking for umbilic points,  i.e.~densities $\{\rho_\mu\} $ for which two or more characteristic velocities $c_\la$ coincide.  We find that for the model  (\ref{eq:rates}) with $K=1$ (two-lane system) no umbilic points (UP) exist in phase space of densities, while for any $K > 1$  a one-parameter UP-containing manifold can be found. 
Namely, UPs appear whenever the average densities on all lanes become all equal, $\rho_\la\equiv \rho$: indeed, analyzing the Jacobian $J_{\mu\la}$
we find an umbilic point of order $K$, with $c_1=c_2 = \ldots =c_K \equiv c_u$, $ c_{K+1} \equiv c_s$, where
\begin{align}
& c_u = 1-2 \rho +  a \rho ( K-1 + \rho  (1-2K)),\\
& c_s = 1-2\rho + K a \rho(2-3 \rho),
\end{align}
and the lower indices $u,s$ in the above refer to an umbilic and a single mode respectively.
The corresponding single mode eigenvector follows from (\ref{eq:Jeiv}) 
\begin{align}
\ket{s} & = (1,1,\ldots  1)^T,
\end{align}
while eigenvectors of the umbilic modes $\ket{1},\ldots, \ket{K}$ form a basis in the  orthogonal complement of $\ket{s}$.
Thus for any number of lanes $K$ we have a line of umbilic points, 
of order $K$ (we henceforth refer to it as an umbilic line),  $\rho_\la\equiv \rho$,  for any value of interlane interaction $a$.

\begin{figure}[ptb]
\center
\includegraphics[width=\linewidth]{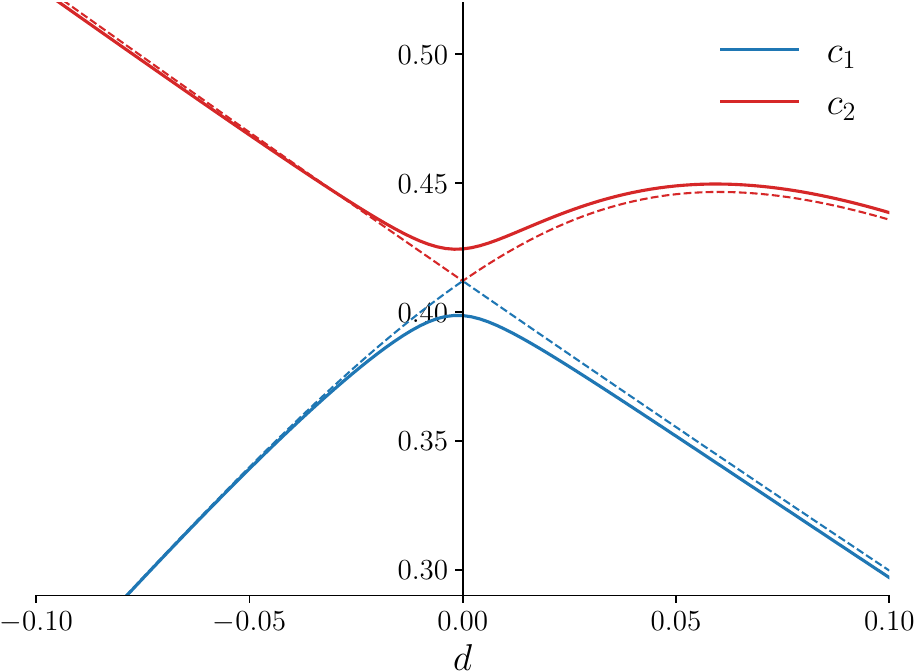}
\caption{Characteristic velocities $c_1,c_2$, stemming from the degenerate umbilic mode,  along  trajectories
$\vec{\rho}(d)$ in phase space
of densities, passing through
(or near) the umbilic line,  on  a manifold orthogonal to it. The parametrization is chosen so that $d=0$ corresponds to the minimal distance to the umbilic line. Dashed curves correspond to a trajectories passing through the UP while for solid curves a small mismatch $\eps=0.005$ is employed. 
Explicitly,  for dashed curves,  $(\rho_1,\rho_2,\rho_3) =(0.3,0.3,0,3) + d(1,1,-2)/\sqrt{6} $ while for solid curves,  $(\rho_1,\rho_2,\rho_3) =(0.3+\eps,0.3-\eps,0.3) + d(1,1,-2)/\sqrt{6}$.
Other parameters: $K=2$, $a=0.4$.  The third characteristic velocity $c_3$ of the nondegenerate mode is not shown. 
The curves show the typical topology of an isolated umbilic point, see also the main text.
}
\label{FigUPisolated}
\end{figure}

\begin{figure}[ptb]
\includegraphics[width=\linewidth]{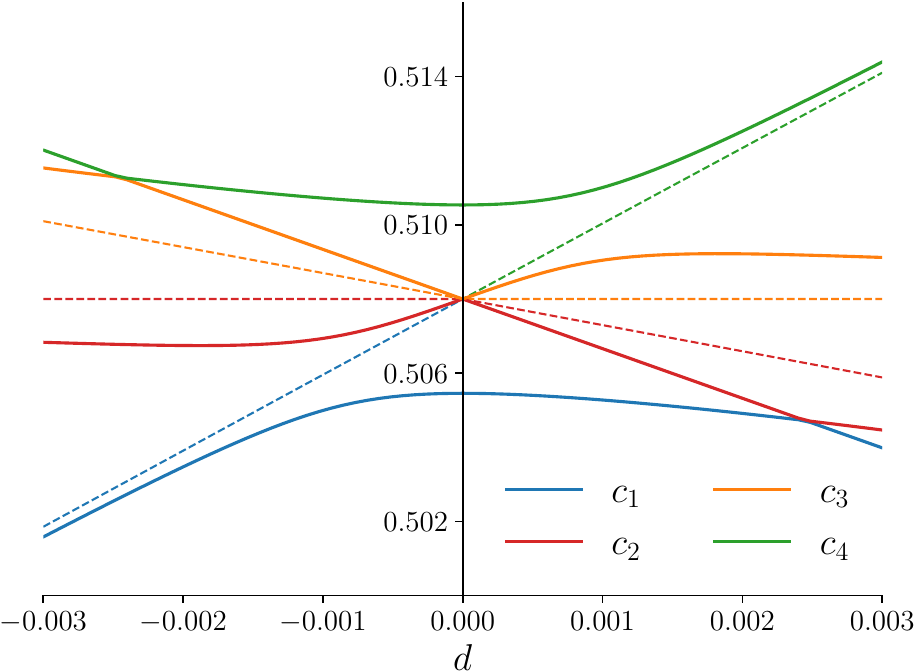}
\caption{Characteristic velocities in the vicinity of an UP with fourfold degeneracy. Two solid curves given by $\vec{\rho}(\eps,d) - (\rho, \rho, \rho, \rho, \rho) = d(1,1,-2,0,0) + \eps (0,0,0,1,-1)$
with $\eps =10^{-3}$ cross the dashed curves $\vec{\rho}(0,s)$ at the origin, indicating a non-isolated UP. Other parameters: $K=4$, $a=0.4$, $\rho=0.3$.
}
\label{FigUPnonisolated}
\end{figure}

Any infinitesimal  shift in the phase space of densities
away from the line of equal densities (umbilic line)
lifts the degeneracy and makes the system strictly hyperbolic, i.e.~all the characteristic velocities $c_\la$  become non-degenerate.  Analyzing the behavior of $c_\la$ in an $\epsilon$-vicinity
of the umbilic line, in any direction orthogonal to the umbilic one, we find for $K=2$ an isolated point UP topology, for any value of interaction $a\neq 0$, see Fig.~\ref{FigUPisolated} for an illustration.  (An umbilic point is \emph{isolated} if there exists an $\epsilon>0$ such that all of the characteristic velocities $c_1,c_2,\ldots c_K$ stemming from the umbilic point $c_1(UP)=\ldots =c_K(UP)=c_u$ are different from $c_u$ in the $\epsilon$-vicinity of the UP).
For $K>2$  umbilic points are no longer isolated, see Fig.~\ref{FigUPnonisolated}.

We are interested in the dynamic structure matrix $\tilde{S}(x,t)$ at the umbilic point $\rho_\la \equiv \rho$ which at large $x,t$ is expected to exhibit universal behavior.  Its components are given by
\begin{equation}
\tilde{S}_{\lambda,\mu}(x,t)=\bra{e_\lambda}\tilde{S}\left(x,t\right)\ket{e_\mu}=\erw{n_{x}^{\lambda}(t)n_{0}^{\mu}(0)}-\rho^2,
\end{equation}
where $(e_\mu)_\nu=\de_{\mu \nu}$. By symmetry we have 
\begin{equation}
\tilde{S}_{\lambda\mu}=\begin{cases}
\tilde{S} & :\lambda=\mu\\
\tilde{S}' & :\lambda\not=\mu
\end{cases} \label{eq:Ssymmetries}
\end{equation}
We normalize the dynamic structure matrix by rewriting it in the basis of  normal eigenmodes, namely by choosing an appropriate transformation $R$,  such that $R J R^{-1}$ is diagonal and $R K R^T=I$ where $K$ is the covariance matrix (\ref{eq:Covariance}). 
An appropriate $R$ is given by the ratio $R =U/\sqrt{\rho(1-\rho)}$, where the rows of the unitary matrix $U$ are normalized eigenvectors of the Jacobian  (\ref{eq:Jeiv}) with eigenvalues ($c_u,c_u, \ldots , c_u, c_s$). Then we 
have
\begin{align}
&R J R^{-1} = diag(c_u,c_u, \ldots , c_u, c_s),\\
&R \tilde{S} R^T \equiv S= diag(S_{u},S_{u},\ldots , S_u, S_s),  \\ 
& \mbox{or} \quad S=S_{u}\left(I-\ket{s}\bra{s}\right) + S_{s}\ket{s}\bra{s},
\label{StructSym}\\
&S_{u}  =\frac{\tilde{S}-\tilde{S}'}{\rho(1-\rho)},\\
&S_{s}  =\frac{\tilde{S}+K \tilde{S}'}{\rho(1-\rho)},
\end{align}
where $S_{u}$ and $S_s$  are normalized dynamical structure factors for umbilic and  single modes respectively.  
Denoting by $\varphi_\la(k) = (n_k^\la(t) - \rho)$ the density fluctuations  on lane $\la$, the stationary fluctuations $\phi_\al$ of transformed  components $\vec{\phi} = R \vec{\varphi}$ in the hydrodynamic limit $\phi_\al(k) \rightarrow \phi_\al(x)$ satisfy canonical relations
\begin{align}
&\erw{\phi_\al(x) \phi_\be(0)} =  \de_{\al \be} \de(x).
\end{align}
The equations of nonlinear fluctuating hydrodynamics for transformed variables at the UP have the form 
\begin{align}
&\partial_t \phi_\la = - \partial_x \left( c_\la \phi_\la +  \bra{\phi} G^\la \ket{\phi} - \partial_x (D \vec{\phi})_\la + (B \vec{\xi})_\la
\right) \label{eq:NLFH}
\end{align}
where $G^\la$ is a so-called mode-coupling matrix, describing the effective couplings of normal mode $\la$ to all other modes and to itself \cite{Spoh14,PSSS2016} while $B \vec{\xi}$ describes noise and $D$ is a phenomenological diffusion matrix. The mode-coupling matrices are obtained in usual way
 \begin{align}
&G^\al= \frac12 \sum_\be \tilde R_{\al \be} (\tilde R^{-1})^T H^\be \tilde R^{-1},
\end{align}
where ${\tilde R} = R/\sqrt{\rho(1-\rho)}$ and
$H^\al$ is the Hessian $(H^\al)_{\be \ga} = \partial^2 j_\al/(\partial \rho_\be \partial \rho_\ga) $.

The long-time evolution of the dynamic structure factors $S_u, S_s$  at large space-time scales is described by 
mode-coupling equations
\begin{align}
&\partial_t S_\alpha(x,t) = (-c_\alpha \partial_x + D_\alpha \partial_x^2) S_\alpha(x,t) \nonumber \\
&+ \int_0^t \rmd t^{\prime} \int_{-\infty}^\infty \rmd x^{\prime}\, S_\alpha(x-y, t-t^{\prime}) \partial_y^2 M_{\alpha \alpha}(y,t^{\prime}),
\label{EQSmode-coupling}\\
&M_{\alpha\alpha}(y,t) = 2 \sum_{\beta,\gamma=1}^{K+1} (G^{\alpha}_{\beta\gamma})^2 S_\beta(y,t)S_\gamma(y,t)\label{MemoryKernel}
\end{align}
where $S_{n\leq K} = S_u$, $S_{K+1} = S_s$, $D_\alpha$ is the diagonal element of the phenomenological diffusion matrix for the eigenmode $\alpha$ and $M_{\alpha\alpha}(x,t)$ is a memory kernel.
Due to symmetry in Eq.~(\ref{StructSym}) the memory kernel can be written as (omitting variable-dependence for brevity) 
\begin{align}
&\frac12 M_{\alpha\alpha}=
\sum_{\be,\ga=1}^K  \left( G^\al_{\be \ga}  \right)^2 S_u^2+\\
&+ \sum_{\be=1}^K  \left( G^\al_{\be, K+1}  \right)^2 S_u S_s +  \left( G^\al_{K+1, K+1}  \right)^2 S_s^2.
\end{align}
Summing the above, we obtain the
memory kernels
\begin{align}
&\ka M_{\alpha\alpha}= (K-1)(g_1)^2 S_u^2+ 2 \left(g_2\right)^2  S_u S_s, \label{eq:Muu}
\\
&\quad \al=1,\ldots ,K \no
\\
&\ka M_{K+1,K+1}= K(g_1)^2 S_u^2+(g_3)^2  S_s^2, \label{eq:Mss} \\
& \ka = \frac{K+1}{2 \rho(1-\rho)}, \no
\end{align}
where $g_\al \equiv g_\al (\rho,a)$ are given by
\begin{align}
&g_{1}(\rho,a)=1+a+a\rho(K-2), \label{eq:g1}\\
&g_{2}(\rho,a)=1+a\tfrac{1-K}{2}+a\rho\left(2K-1\right),\label{eq:g2}\\
&g_{3}(\rho,a)=1+aK\left(3\rho-1\right).\label{eq:g3}
\end{align}
Since,  due to a symmetry,  memory kernels for all umbilic modes $\al = 1,2, \ldots ,K$ are equal,  the system (\ref{EQSmode-coupling}) contains just two distinct equations,  one for the umbilic mode $S_u$ with degeneracy $K$ and another for the single mode $S_s$.  
The system of two equations can be written in terms of effective $2\times 2$ mode coupling matrices 
$\bar G^u \equiv G^1$,  $\bar G^s \equiv G^{K+1}$
\begin{align}
&\bar{G}^{u} \equiv \bar{G}^{\alpha \leq K} = \sqrt{\tfrac{\rho(1-\rho)}{K+1}} \left(\begin{array}{cc}
g_{1}\sqrt{K-1} & g_{2}\\
g_{2} & 0
\end{array}\right),   \label{GuEff} \\
&\bar{G}^{s} \equiv \bar{G}^{K+1} = \sqrt{\tfrac{\rho(1-\rho)}{K+1}} \left(\begin{array}{cc}
g_{1}\sqrt{K} & 0\\
0 & g_{3}
\end{array}\right).   \label{GsEff}
\end{align}

Here we need to stress that  (\ref{GuEff}) and (\ref{GsEff}) are 
\textit{effective} mode-coupling matrices and respectively (\ref{EQSmode-coupling})
are \textit{effective} mode-coupling equations, based on symmetries
(\ref{eq:Ssymmetries}) occurring at the umbilic manifold. In the effective MCT (\ref{EQSmode-coupling}),(\ref{GuEff})  the umbilic mode appears as a non-degenerate mode with self-coupling proportional to $g_1$.  
We emphasize that we cannot  expect (\ref{EQSmode-coupling})  to describe the umbilic mode quantitatively; for the latter purpose we need full system dynamics, captured by e.g.~full Monte-Carlo simulations.

Nevertheless, it is instructive to
test the predictions of mode-coupling theory for the two effective modes, the umbilic and single mode. 
According to the scaling hypothesis, $S_\alpha(x,t) $ at large $x,t$
has universal scaling behavior
\begin{equation}
S_{\alpha}(x,t) \simeq (E_\alpha t)^{-1/z_\alpha} f_{\alpha} \left((x-c_\alpha t)(E_\alpha t)^{-1/z_\alpha}\right),
\label{{scalingform}}
\end{equation}
where $f_{\alpha}$ is a scaling function, $z_\alpha$ is a dynamical exponent and $E_\alpha$ is a non-universal scaling factor.
First, note that  from non-negativity of the rates  (\ref{eq:rates}) we have $-1/K \leq a$ and  from (\ref{eq:g1}) it follows that 
\begin{align}
&\min_{\rho,a} g_{1}(\rho,a)= \min_\rho \left(1-\frac{1+\rho (K-2)}{K}\right) = \frac{1}{K} >0
\end{align}
is always positive.  Then the degenerate mode (\ref{GuEff}) is a mode with self-coupling, $\bar{G}^{u}_{11}\sim g_1 >0$. Power counting based on mode-coupling equations yields the standard  prediction  
\begin{align}
&z_u = \frac{3}{2} \label{eq:zKPZ}
\end{align} 
for the dynamical exponent of the umbilic mode.  
For the remaining single mode,  accounting for $\bar{G}^{s}_{11} \sim g_1 >0$, two scenarios are possible \cite{JSP2015}:
(A) if $\bar{G}^{s}_{22} \neq 0$  then the single mode is a KPZ mode $z_s = 3/2 $ and $S_s\equiv f_{\rm KPZ}$ \footnote{Note that the mode-coupling prediction for the scaling function of a single-mode with self-coupling \cite{Colaiori01} is distinct from $f_{\rm KPZ}$ but is known to be incorrect \cite{Prae04}, with the latter being the correct scaling function in this scenario.}.
(B) if $\bar{G}^{s}_{22} \sim g_3 = 0$, i.e. 
\begin{align}
&1+aK\left(3\rho-1\right)=0,  \label{eq:HeatModeCondition}
\end{align} 
  then the  Fibonacci universality class  \cite{2015Fibonacci,JSP2015} criterium predicts 
\begin{align}
&z_s = 1+ \frac{1}{z_u}= \frac{5}{3} \label{eq:zHeat}
\end{align} 
with, additionally, the explicit scaling function $f_s$
\begin{align}
f_s(r) &= 
\int_{\mathbb{R}}e^{- E_s |p|^{z_s}\left[1-i\cdot\mathrm{tan}\left(\frac{ \pi z_s}{2}\right)\mathrm{sgn}\left(p(c_s - c_u)\right)\right]}e^{ipr}
\frac{\mathrm{d}p}{2\pi}
\label{5/3LeviStable}
\end{align}
which is a maximally  asymmetric Levy stable $ \frac{5}{3}$ 
distribution. The only free parameter in (\ref{5/3LeviStable}) is a nonuniversal  scaling factor $E_s$, which cannot be predicted from mode-coupling theory \cite{PSSS2016,BeiSS2020}.
For the umbilic mode we determine the scaling factor as \cite{Prae04}
\begin{equation}
    E_u = 2^{3/2}|G^u_{uu}| \label{Eu_eq}
\end{equation}
 while for the heat mode we determine the scaling factor $E_s$ by matching the maxima of the scaling function \eqref{5/3LeviStable}.

\section{Simulation results}
\begin{figure}[ptb]
\center
\includegraphics[width=0.8\linewidth]{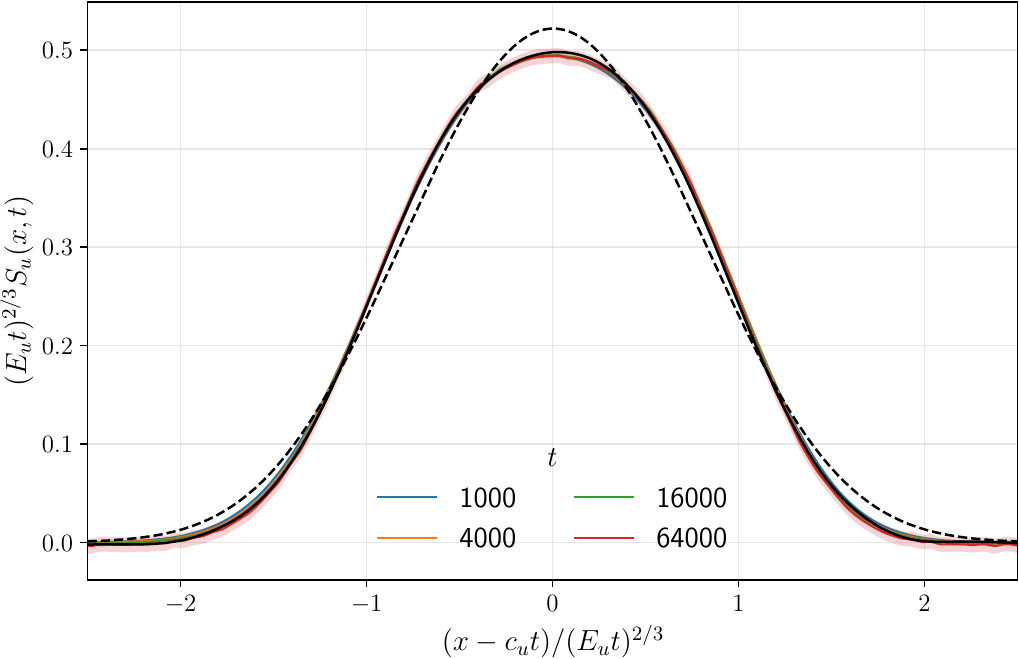}
\caption{
Data collapse for the doubly degenerate (K=2) umbilic mode from lattice Monte Carlo simulations with $z_u = 3/2$, $c_u = 5/8$.  Shaded regions show two standard deviation neighborhoods. Full black curve, overlapping with our data, shows the (scaled) $S_{11}$  correlator of Ref.~\cite{2025Spohn} at $X=Y=-1$. Black dashed curve shows the best-fit KPZ (Pr\"ahofer-Spohn) scaling function. 
Simulation parameters: $K=2$, $a=2$, $\rho = 1/4$, system size $N = 10^6$, $n_{\textrm{samples}} = 10^3$.}
\label{FigUmbilicLattice}
\end{figure}

\begin{figure}[ptb]
\center
\includegraphics[width=0.8\linewidth]{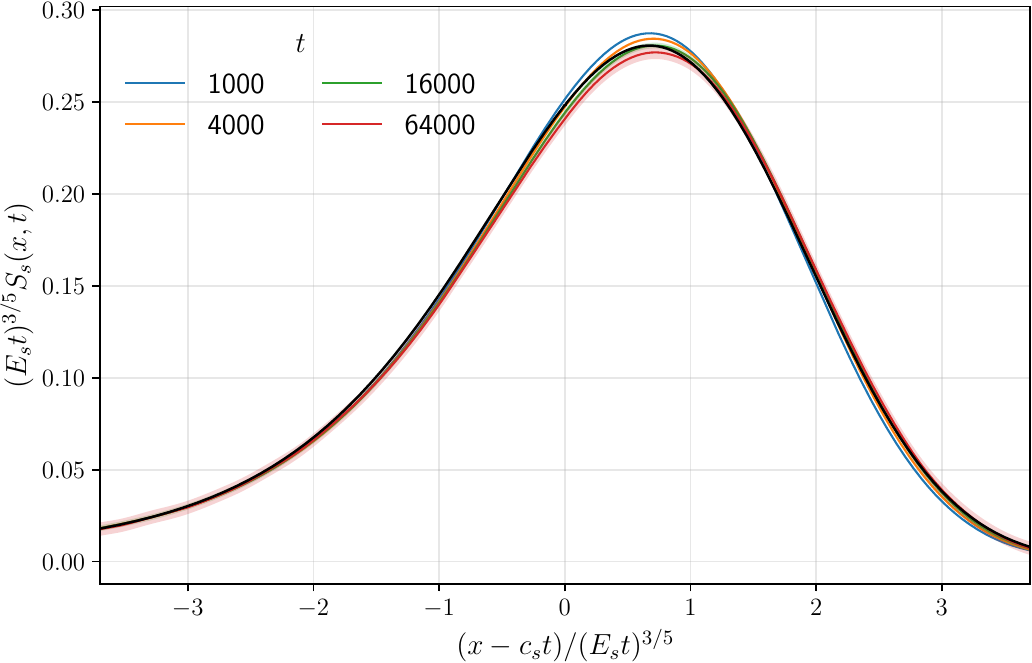}
\caption{
Data collapse for the heat mode from lattice Monte Carlo simulations with $z_s = 5/3$, $c_s = 7/4$ and $E_s = 1.30$. Shaded regions show two standard deviation neighborhoods. Best collapse of the data is obtained for $z_s \approx 1.645$. Black curve shows the maximally  asymmetric Levy stable $\frac{5}{3}$ distribution \eqref{5/3LeviStable}.
Simulation parameters as in Fig.~\ref{FigUmbilicLattice}.}
\label{FigHeatLattice}
\end{figure}

As stressed after Eq.(\ref{GsEff}), only considering the full dynamics of the system can shed light on the qualitative features on our model, especially for the degenerate umbilic mode.
We have performed large-scale Monte-Carlo simulations of the model on Fig.~\ref{FigMultichainTASEP}  on the umbilic line to test the mode-coupling theory (MCT) predictions. We concentrate on  the case (B),  i.e.~$g_3 =0$ in (\ref{eq:g3}), which is the most interesting scenario. Our results fully confirmed the validity of the MCT predictions ~Eqs.(\ref{eq:zKPZ}), (\ref{eq:zHeat}), and (\ref{5/3LeviStable}), see Figs~\ref{FigUmbilicLattice},\ref{FigHeatLattice}.  

For the umbilic mode with double degeneracy $K=2$ we find an evidence of the dynamic exponent $z_u=3/2$ (\ref{eq:zKPZ}); however the umbilic scaling function $f_u$  appears very different from the Prahofer-Spohn (KPZ universality) scaling function. Moreover, $f_u$ differs from  any known scaling function, which can be attributed to $z=3/2$ and is symmetric: not only KPZ,  but also the e.g.~Moore-Colaiori function \cite{Colaiori01},  or symmetric Levy stable $3/2$ distribution. 
From our numerical data, we can tabulate $f_u$ for a doubly degenerate umbilic mode $K=2$ with relatively good precision. Remarkably, the shape of the umbilic function is robust to changes of system parameters, i.e.~appears universal on the whole umbilic line, see Fig.~\ref{FigHeat_a}. A possible reason for this universality may be the  following: 
curiously, our umbilic mode considered alone, i.e.~ignoring the nondegenerate mode, falls into the umbilic scenario of $2$-component
systems described in Ref.~\cite{2025Spohn}, and 
can be  classified according to it as a special point on a fixed points line,
see \ref{app:E} for details. Further confirmation of this reasoning is that, within our numerical precision, the scaling function from continuous time Burgers equation analysis in Ref.~\cite{2025Spohn} match our umbilic scaling function determined from Monte-Carlo simulations, see Fig~\ref{FigUmbilicLattice}.

As far as the nondegenerate mode is concerned,  our findings agree well with mode-coupling theory predictions (\ref{eq:zHeat}), (\ref{5/3LeviStable}), see Fig.~\ref{FigHeatLattice}. 
Note that due to absence of the off-diagonal terms in $\bar{G}^{s}$ from (\ref{GsEff}) finite-time effects are small and the Levy stable $5/3$ distribution (\ref{5/3LeviStable}) is established already for relatively small times.
Within our numerical accuracy, starting from times of order $t =10^4$  Monte Carlo steps we see no deviations from the analytical result (\ref{5/3LeviStable}) apart from a small deviation in the finite-time dynamical exponent.

Our findings for the single mode do not depend on the value of interaction $a$, as long as the (B) condition $g_3=0$ (\ref{eq:HeatModeCondition}) is satisfied.  

\clearpage

\bigskip 
\subsection{Umbilic universality} 
\begin{figure}[ptb]
\scalebox{0.48}{\includegraphics{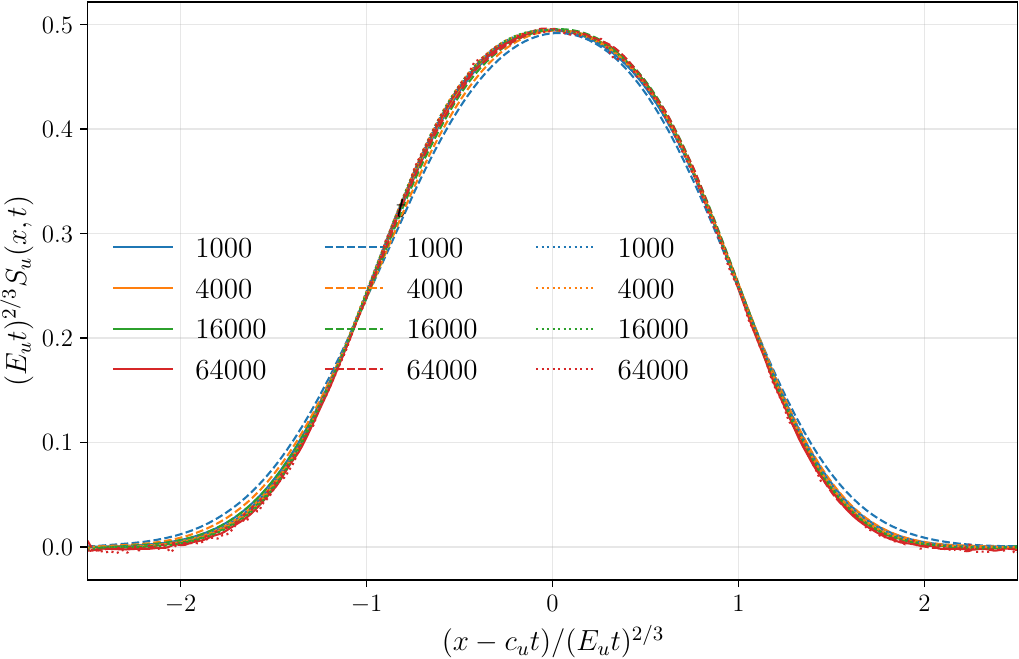}}
\scalebox{0.48}{\includegraphics{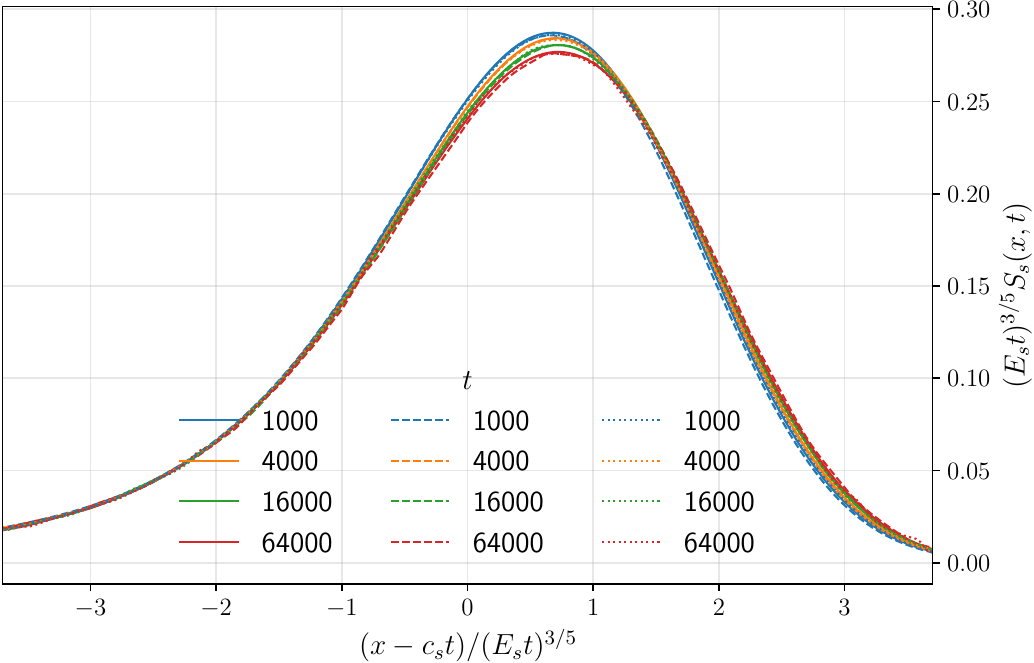}}
\caption{
Comparison of scaled umbilic (left panel) and heat (right panel) modes from lattice Monte Carlo simulation for $a=2$, $\rho=1/4$ (full curves), $a=1$, $\rho=1/6$ (dashed curves) and $a=3$, $\rho=5/18$ (dotted curves), satisfying the condition (\ref{eq:HeatModeCondition}) for the absence of self-coupling. Other simulations parameters as in Fig.~\ref{FigUmbilicLattice}.
}
\label{FigHeat_a}
\end{figure}

We find that the umbilic scaling function  does not depend on the interlane interaction parameter for fixed $K$, see Fig.~\ref{FigHeat_a}.
In this respect the scaling function appears  more robust than the one of a two-lane bidirectional particle  model \cite{2024Spohn, 2025UmbilicBidirectional} where the umbilic scaling function varies the interlane interaction parameter, called $\ga$ in \cite{2024Spohn, 2025UmbilicBidirectional}.

\section{Numerical integration of continuous NLFH equations}
\begin{figure}[ptb]
\scalebox{0.48}{\includegraphics{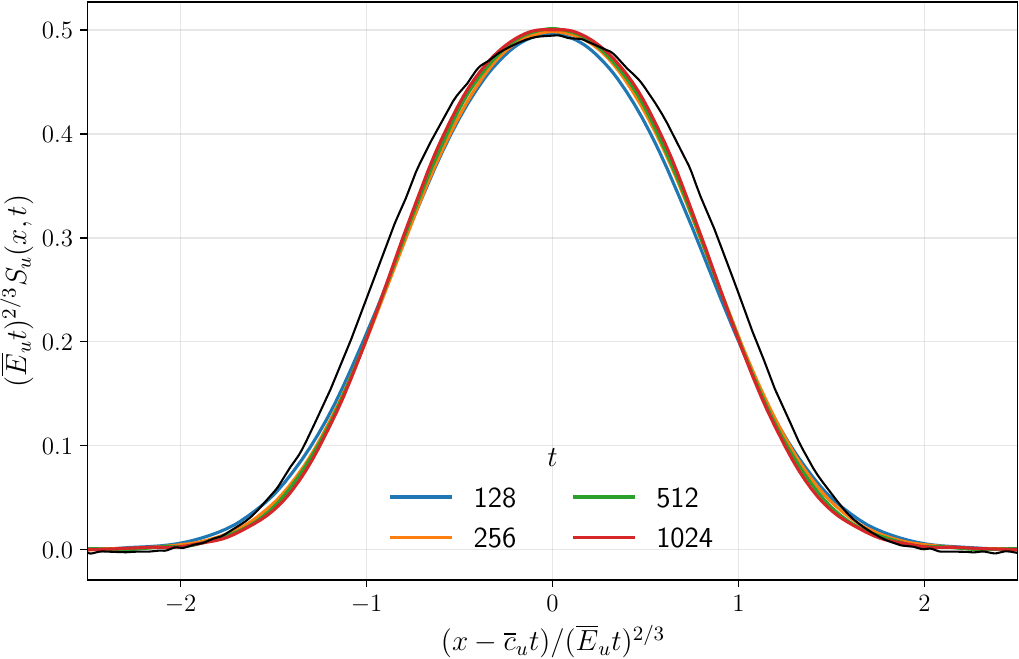}}
\scalebox{0.48}{\includegraphics{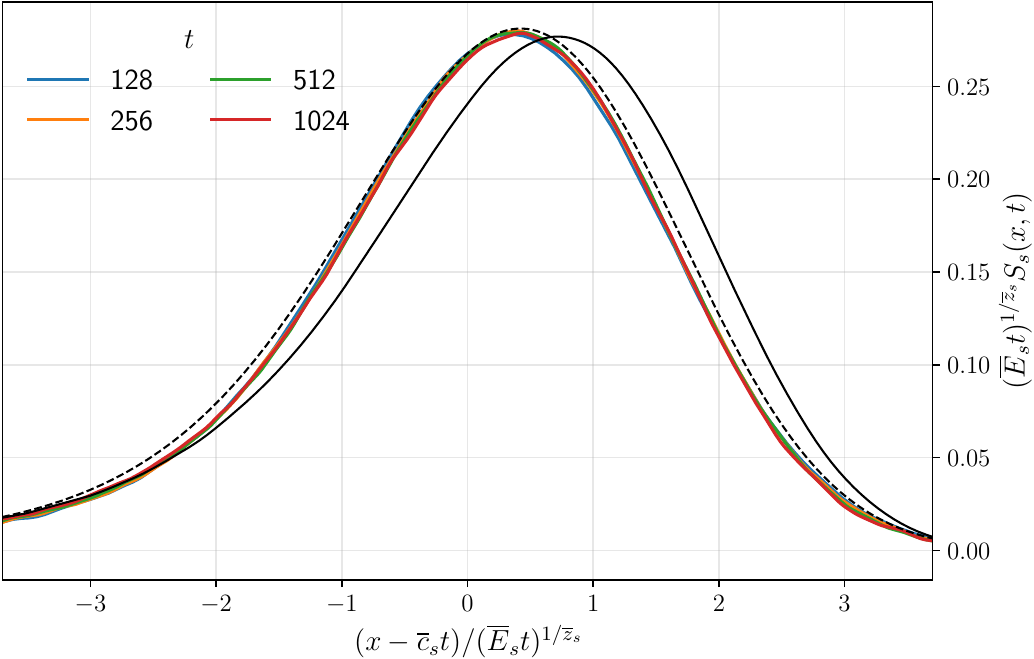}}
\caption{
Data collapse from continuous NLFH simulation for:
(left panel) the umbilic mode  with $\overline z_u = 3/2$, $\overline c_u=0.575$ and $\overline E_u = 2.32$ with $\overline E_u$ determined by matching the height of the umbilic scaling function estimated from Monte Carlo simulations (black curve).
(right panel) the for heat mode where the best data collapse is obtained for $\overline z_u = 1.74$, $\overline c_u=1.75$ and $\overline E_s = 0.94$.
Black curves show maximally asymmetric Levy stable $\frac{5}{3}$ (full line) and $\overline z_s$ (dashed line) distributions. 
Simulation parameters: $K=2$,  $\rho=1/4$, $a=2$ with effective mode-coupling matrices
$\bar{G}^{u}=\frac{3}{8}\left(\begin{matrix}2 & 1\\ 1 & 0\end{matrix}\right)$ and
$\bar{G}^{s}=\frac{3}{8}\left(\begin{matrix}2\sqrt{2} & 0\\ 0 & 0 \end{matrix}\right)$ and discretization parameters $\Delta t = 2\times 10^{-3}$, $L= 2^{20}$.
}
\label{FigUmbilicHeatBurgers}
\end{figure}

 To check our findings in the lattice model we have also performed an numerical analysis of the stochastic system of equations of nonlinear fluctuating hydrodynamics stemming from the original hydrodynamic equations (\ref{eq:continuity_eq}),  namely, 
\begin{align}
&\partial_t \vec{u} = - \partial_x \left( J  \vec{u}  +  \frac12 \bra{u} \vec{H} \ket{u} - \partial_x (D \vec{u})+ (B \vec{\xi})
\right) \label{eq:OriginalSystem}
\end{align}
where $J$ is the Jacobian, $\vec{H}$ is the Hessian  $(H^\al)_{\be \ga} = \partial^2 j_\al / (  \partial \rho_\be \partial \rho_\ga) $ 
and $u_\al(x,t)$ is the deviation of the respective density.  Using (\ref{eq:current-densityRelation}), on the diagonal $\rho_j \equiv \rho$ we have 
\begin{align}
&J_{\al \be} = (1-2\rho) (1+Ka \rho )\, \de_{\al \be} + a \rho (1-\rho) (1-\de_{\al \be}),
\label{eq:Jac}\\
&(H^\al)_{\be \ga} =-2 (1+Ka \rho ) \,\de_{\al \be} \de_{\be \ga}, \no\\
& + a  (1-2\rho) (1-\de_{\be \ga}) (\de_{\al \be} +\de_{\al \ga}  )
\label{eq:Hessian}
\end{align} 
For $K=2$ (three lanes) we have a set of $K+1=3$ coupled equation,  of the form  
\begin{align}
&\dot u_1 = -\partial_x [
d u_1 + e(u_2+ u_3) + \frac12 (f u_1^2 + 2g(u_2+u_3) u_1  )  \no \\
&+D_1 \partial_x u_1  + B \vec{\xi_1}],  \quad \mbox{etc.}
\end{align}
(the Eqs for $\dot u_2, \dot u_3$ are written analogously),
where we denoted
\begin{align}
& g= a  (1-2\rho) ,\quad  e=a \rho (1-\rho) \no\\
&f =-2 (1+Ka \rho ),\quad  d = (1-2\rho) (1+K a \rho ), \quad K=2.
\label{eq:defs}
\end{align}
and the covariance matrix is diagonal, 
\begin{align}
  &  \langle u_\al(x),u_\be(x')\rangle=\rho (1-\rho) \delta_{\al \be} \delta(x-x'). \label{eq:covarianceOriginal}
\end{align}
Note that we preferred to integrate the original system of equations rather then the same system after transforming to normal modes (i.e.~after diagonalizing the Jacobian) to avoid the ambiguity of the transformation and retain the original symmetry the dynamics. To compare with the Monte-Carlo
data if Figs.~\ref{FigUmbilicLattice} and \ref{FigHeatLattice} we perform the diagonalization to normal modes directly on numerically obtained two-point functions. The results are presented in Fig.~\ref{FigUmbilicHeatBurgers}.

We observe a good agreement in the scaling behavior and shape of the umbilic mode two-point function between lattice Monte Carlo and 
continuous stochastic PDE simulations. For the heat mode, we find a slight disagreement in scaling (expected scaling $z_s=5/3$ in Monte-Carlo simulations, and a slightly different scaling from stochastic PDE simulations), which we attribute to worse convergence properties of the continuous compared to lattice dynamics and non-stationarity of the employed discretization.

\begin{figure}[ptb]
\scalebox{0.48}{\includegraphics{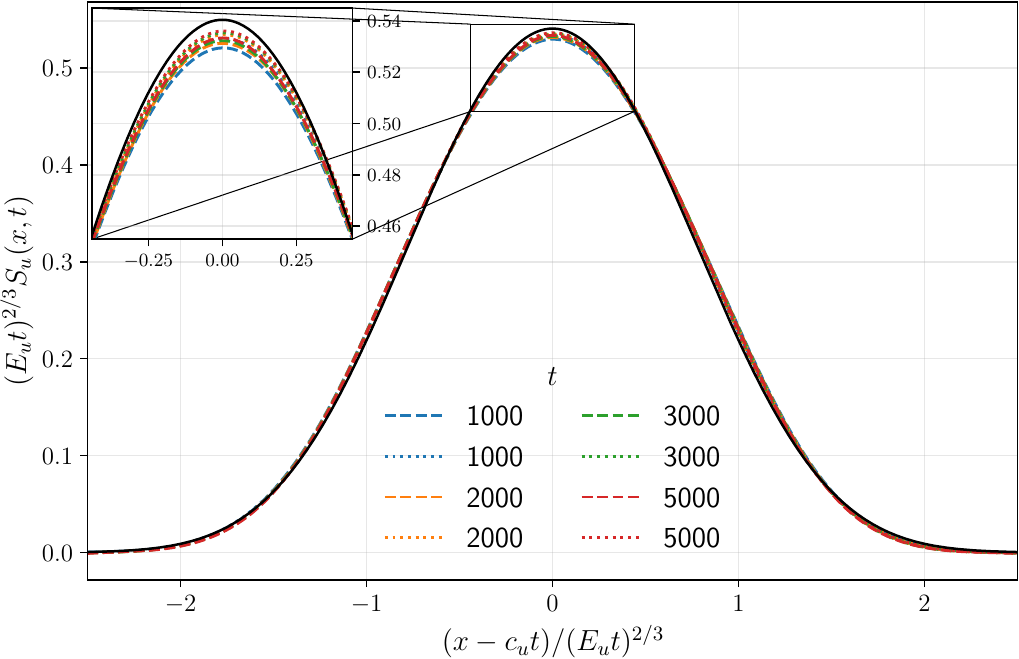}}
\scalebox{0.48}{\includegraphics{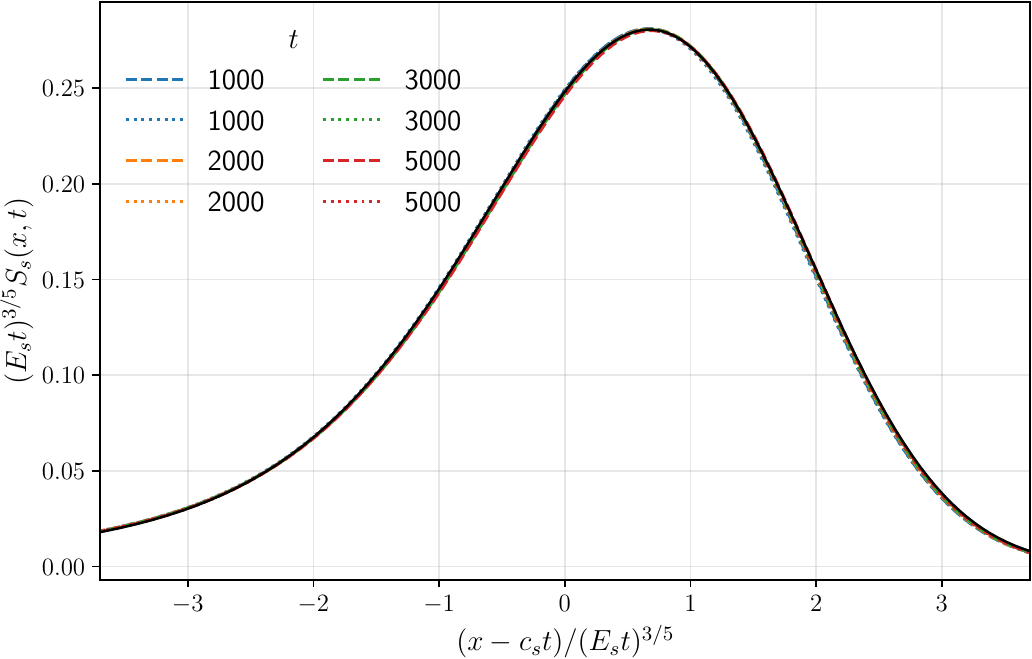}}
\caption{
Data collapse for systems with high (triple $K=3$ (dashed lines) and quadruple $K=4$ (dotted lines)) 
degeneracy, from lattice Monte Carlo simulations with $4$ lanes and $5$ lanes respectively. Collapse for the umbilic mode (left panel) with the best-fit KPZ (Pr\"ahofer-Spohn) scaling function (black line) for comparison and the heat mode (right panel) with the $5/3$ Levy stable mode of maximal asymmetry \eqref{5/3LeviStable} (black line), predicted by mode coupling equations for both $K=3$ and $K=4$ for comparison.
Parameters for $K=3$: $\rho = 1/4$, $a=4/3$,  $E_s = 1.16$  and for $K=4$: $\rho=1/4$, $a=1$, $c_s = 7/4$, $E_s = 1.11$, $c_u = 0.8125$, satisfying the condition (\ref{eq:HeatModeCondition}) for the absence of self-coupling. Two standard deviation neighborhoods are of the order of the line widths and are not shown for clarity.
Other simulation parameters as in Fig.~\ref{FigUmbilicLattice}.}
\label{Fig_K34}
\end{figure}

\begin{figure}[ptb]
\centering
\includegraphics[width=0.8\linewidth]{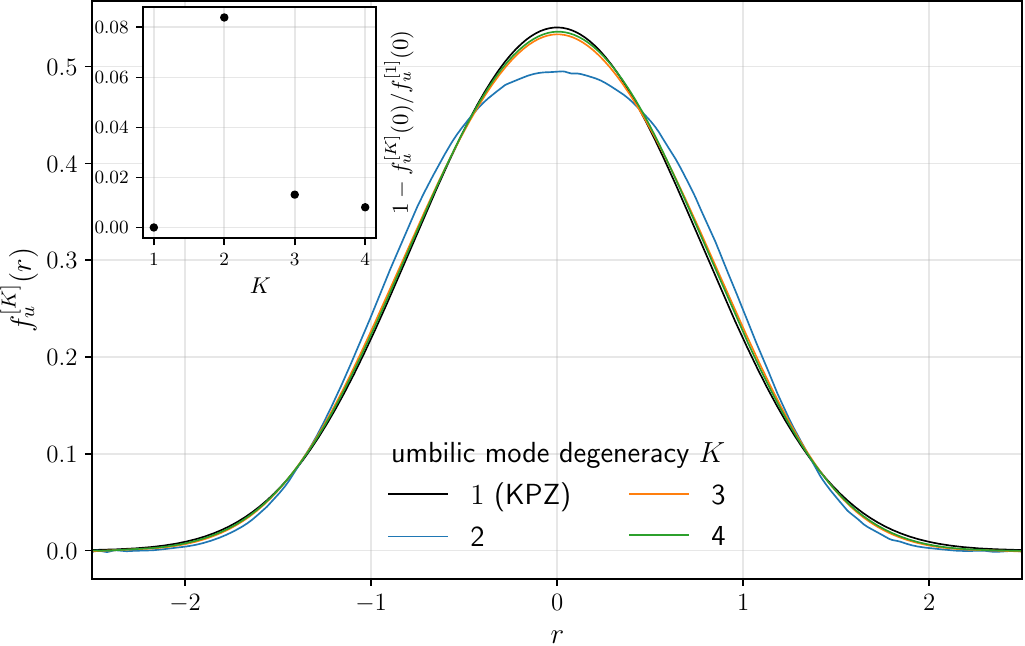}
\caption{
Comparison of umbiling scaling functions $f_{u}^{[K]}$ at different degeneracies $K$ (colored lines) with the best-fit KPZ (Pr\"ahofer-Spohn) scaling function (black line). We observe rapid convergence towards the latter with increasing degeneracy $K$.
Scaling functions estimated from rescaled dynamical strucure factors as $f_{u}^{[K]}(r) \approx (E_u t_{K})^{2/3}S_u(c_u t_K + r(E_u t_{K})^{2/3},t_{K}) $ at times $t_2 = 16000$ and $t_3 = t_4 = 5000$. Inset shows difference of scaling functions heights (values at $r=0$) relative to the KPZ scaling function ($K=1$) as a function of the umbilic degeneracy $K$.
Simulation parameters as in Fig.~\ref{Fig_K34}.}
\label{Fig_K234}
\end{figure}

\section{$K>2$ case: Universality of umbilic points with higher degeneracy. }
Up to now we only described our numerical results for the double-degenerate $K=2$ umbilic mode, while our effective MCT theory predictions, namely Eqs.~(\ref{eq:zKPZ}), (\ref{eq:zHeat}), and (\ref{5/3LeviStable})
are expected to be valid for an umbilic mode of arbitrary
degeneracy $K$.
To test this prediction we have performed a similar analysis for $K=3$ and $K=4$ (see Fig.~\ref{Fig_K34}) which is more challenging numerically, when it comes to Monte-Carlo simulations, but has better finite-time convergence properties.
The predictions for the dynamical exponents $z_s = 5/3$ (\ref{eq:zHeat}) and the scaling function (\ref{5/3LeviStable}) of the heat mode are found to be in excellent agreement with our numerical simulations, see right panel of Fig.~\ref{Fig_K34}.
While the $z_u=3/2$ prediction of MCT (\ref{eq:zKPZ}) remains robust, the umbilic scaling function $f_u$ changes appreciably with the  umbilic mode degeneracy $K$ as shown in the left panel of Fig.~\ref{Fig_K34}, cf.~umbilic scaling function for $K=2$ in Fig.~\ref{FigUmbilicLattice}.
With increasing $K$ the umbilic scaling functions $f_{u}^{[K]}$ starts to closely approach the KPZ scaling function, see Fig.~\ref{Fig_K234}. It would be interesting to investigate if the convergence $f_{u}^{[K]} \rightarrow f_{\rm KPZ} = f_{u}^{[1]}$ takes place in the ``meanfield" limit $K \rightarrow \infty$.

\section{Conclusions}
We have investigated a family of fully asymmetric interacting exclusion processes evolving in parallel on $K+1$ lanes,  with Bernoulli stationary measure. 
For $K\geq 2$ we establish the existence of an umbilic manifold, i.e.~a manifold where all but one characteristic velocities coincide. In contrast with another recently investigated umbilic mode scenarios \cite{2024Spohn, 2025Spohn, 2025SpohnSingle, 2025UmbilicBidirectional},
here, alongside with the degenerate umbilic mode, a usual nondegenerate long-living mode exist,  coupled to the umbilic mode.  
Concentrating on the umbilic manifold, we  investigated the asymptotic behavior of the umbilic mode, which is expected to have universal behavior.  We carried out both  Monte-Carlo simulations of a particle system on a lattice and  simulations of  the respective continuity/Burgers-type  equations with diffusion and stochastic noise. We found an evidence of a family of universal umbilic structure factors,  appearing in asymptotic space-time behavior of a long-lived hydrodynamic modes with equal characteristic velocities  (umbilic modes). The umbilic scaling function appears to be symmetric,    
characterized by the ``KPZ" dynamical scaling exponent $z=3/2$,  but with a shape distinctly different from the Prahofer-Spohn scaling function  \cite{Prae04}.  The shape of umbilic function depends on the umbilic degeneracy. We derived effective mode-coupling equations,
which describe well some features of our multilane model: in particular, 
they give a perfect analytic description for the nondegenerate mode, coexisting and interacting with the umbilic mode, for arbitrary degeneracy.  On the other hand,  an analytic description of the family of umbilic scaling functions at fixed degeneracy, generalizing the Prahofer-Spohn result \cite{Prae04} appears a challenging open problem. 

\paragraph*{\bf Acknowledgements}
\v{Z}.K. is supported by the Simons Foundation as a Junior Fellow of the Simons Society of Fellows (1141511).
V.P.  acknowledges support by ERC Advanced grant No.~101096208 -- QUEST, and Deutsche Forschungsgemeinschaft through DFG project KL645/20-2. We thank Herbert Spohn and Dipankar Roy for providing the raw scaling function data of Ref.~\cite{2025Spohn}.\\

\appendix

\section{\\Monte-Carlo simulations of a multichain model: dynamic structure factor}
\label{app:B}

Initial states are drawn from the factorized stationary distribution of the process. The two-point function is estimated using translation invariance and stationarity, which allow for the computation of the spatial and temporal averages leading to the Monte-Carlo estimator
\begin{align}
\tilde\sigma_{x,t}^{\lambda \mu}(M,\tau,L) =&
\frac{1}{LMR} \sum_{l=1}^L \sum_{j=1}^M
n_{l+x, j\tau+t}^{\lambda,(r)}n_{l,j\tau}^{\mu,(r)}.
\label{eq-sigma}    
\end{align}
Finally, $\tilde S_{\lambda \mu}(x,t)$ is obtained by averaging over $P=n_{\rm sample}$ independently generated and propagated initial configurations
of $\tilde\sigma_{L,x}^{\lambda \mu}$, i.e.
\begin{align}
\tilde{S}_{\lambda\mu}(x,t)=
\frac{1}{P}\sum_{p=1}^P \tilde\sigma_{L,x}^{\lambda \mu, (p)}
- \rho_\lambda\rho_\mu
+\mathcal{O}(P^{-1/2}).
\end{align}
The error estimates for
$\tilde S_{\lambda \mu}(x, t)$ are calculated from $P$ independent measurements, whereas $L$, $M$, $\tau$ are variance reduction parameter.
\section{\\Numerical simulations of multilane Burgers equations}
\label{app:C}
We seek a numerical scheme that efficiently approximates the solutions of the following system of Burgers equations
\begin{align}
&\partial_t \phi_\la = \partial_x \left( c_\la \phi_\la +   \sum_{\alpha, \beta}G^\la_{\alpha \beta} \phi_\alpha \phi_\beta + D_\lambda \partial_x \phi_\la + B_\lambda \xi_\lambda
\right). \label{app:NLFH}
\end{align}
We approximate the continuous fields $\phi_\lambda$ by discrete points $\phi^{\lambda}_{x, t}$ where $x \in \mathbb{Z}_L$, $t \in \mathbb{Z}$ with spacing 1 and $\Delta t$ in the spatial and temporal directions respectively. We impose periodic boundary condition by identifying lattices indices as $L+1 \equiv 1$. Along the spatial direction we use the discretization of Ref.~\cite{2024Spohn} while along the temporal direction we adopt a semi-implicit discretization motivated by Appendix A of Ref.~\cite{Hairer_Voss_2011}. The resulting set of equations reads
\begin{equation}
    \sum_y T^\lambda_{x,y} \phi^\lambda_{y, t+1} = \phi^\lambda_{x, t} +   \Delta t\, \mathcal{L}^\lambda_{x, t} + \Delta t\, \mathcal{N}^{\lambda}_{x, t} + (\Delta t)^{1/2} B_\lambda(\phi^\lambda_{x, t} - \phi^\lambda_{x-1, t}) \label{discrete_system}
\end{equation}
where $T$ is a cyclic tridiagonal matrix ($\delta$-function indices are taken modulo $L$)
\begin{equation}
T^\lambda_{x,y} = (1 + D_\lambda \Delta t) \delta_{x, y} - \tfrac{D_\lambda \Delta t}{2}\left(\delta_{x, y+1} + \delta_{x, y-1}\right), \label{cyclic_tridiagonal}
\end{equation}
while $\mathcal{L}$ and $\mathcal{N}$ are the linear and nonlinear parts of the discretization respectively
\begin{align}
    \mathcal{L}^\lambda_{x, t} &= c_\lambda\left( \phi^\lambda_{x,t} -\phi^\lambda_{x-1,t} \right) + \tfrac{D_\lambda}{2}\left(\phi^\lambda_{x+1,t} -2 \phi^\lambda_{x,t} +\phi^\lambda_{x-1,t}\right), \label{linear_disc}\\
    \mathcal{N}^\lambda_{x, t} &= \sum_{\alpha, \beta} \tfrac{G^\lambda_{\alpha\beta}}{3}
    \left( 
-\phi^\al_{x-1,t} \phi^\beta_{x-1,t} - \phi^\alpha_{x-1,t} \phi^\beta_{x,t} + \phi^\alpha_{x,t} \phi^\beta_{x+1,t} + \phi^\al_{x+1,t}\phi^\beta_{x+1,t}
    \right). \label{nonlinear_disc}
\end{align}
The system \eqref{discrete_system} can be solved efficiently due to the cyclic tridiagonal structure of $T$. In particular, using the Sherman–Morrison formula, solving the system \eqref{discrete_system} amounts to solving solving a pair of tridiagonal systems via the Thomas algorithm.

While the linear part of the the spatial discretization (\ref{linear_disc}) exactly preserves the dynamics's stationary measure, we emphasize that the nonlinear part of the discretization (\ref{nonlinear_disc}) does not. The effects of non-stationarity are suppressed for small time steps $\Delta t$ and we have checked that for used time steps the deviations from stationarity at longest times, as measured by the static covariance matrix (\ref{eq:covarianceOriginal}), are at most a few percent. We note that devising a discretization of the nonlinearity that exactly preserves the stationary measure would facilitate more detailed simulations of the NLFH equations.

\section{\\Effective NLFH equations for umbilic mode and comparison with Ref.~\cite{2025Spohn}}
\label{app:E}

The NLFH for  transformed  variables  $\vec{\phi} =  R \vec{\varphi}$ at UP have form, 
\begin{align}
&\partial_t \phi_\la = - \partial_x \left( c_\la \phi_\la +  \bra{\phi} G^\la \ket{\phi} - \partial_x (D \vec{\phi})_\la + (B \vec{\xi})_\la
\right) \label{eq:NLFH}
\end{align}
Because of the degeneracy of the umbilic mode,  the diagonalizing transformation  $R$ has a one-parameter freedom.
Using a \textit{particular} tranformation $R$ of the form 
\begin{align}
&R= (\rho(1-\rho))^{-\frac12} \left(
\begin{array}{ccc}
 \frac{1}{\sqrt{2}} & -\frac{1}{\sqrt{2}} & 0 \\
 \frac{1}{\sqrt{6}} & \frac{1}{\sqrt{6}} & -\sqrt{\frac{2}{3}} \\
 \frac{1}{\sqrt{3}} & \frac{1}{\sqrt{3}} & \frac{1}{\sqrt{3}} \\
\end{array}
\right), \label{app:R}
\end{align}
we obtain the NLFH equations
\begin{align}
&\partial_t \phi_1 = -\partial_x (c_u   \phi_1 + 2b_1  \phi_1 \phi_2 +2b_2  \phi_1 \phi_3  + D_1 \partial_x \phi_1  + B_1 \xi_1 )\\
&\partial_t \phi_2= -\partial_x ( c_u   \phi_2+b_1  (\phi_1^2 - \phi_2^2) + 2 b_2  \phi_2 \phi_3 +D_2 \partial_x \phi_2  + B_2 \xi_2)\label{eq:phi2}\\
&\partial_t \phi_3= -\partial_x (c_v   \phi_3+  b_1 \sqrt{2} (\phi_1^2 + \phi_2^2) +  b_3 \phi_3^2 +D_3 \partial_x \phi_3  + B_3 \xi_3) \\
& \langle \phi_\al(x), \phi_\be(x')\rangle=\delta_{\al \be} \delta(x-x') \label{eq:covarianceStatic}\\
&b_1=G^{1}_{12}=G^{2}_{21}= -\sqrt{\frac {\rho(1-\rho)}{3}} \ \frac{1+a}{\sqrt{2}} \\
&b_2=G^{1}_{13}=G^{1}_{31}=- \sqrt{\frac {\rho(1-\rho)}{3}} \  \left(1+a (3 \rho - \frac12)\right) \\
&b_3=G^{3}_{33} =-  \sqrt{\frac {\rho(1-\rho)}{3}} \  \left(1+2a(3 \rho-1) \right)
\end{align}

The time-asymptotic form of the umbilic part of the system (\ref{eq:phi2}) is
\begin{align}
&\partial_t \phi_1 = -\partial_x (c_u   \phi_1 + 2b_1  \phi_1 \phi_2  + D_1 \partial_x \phi_1  + B_1 \xi_1 ) \label{eq:phi1-B} \\
&\partial_t \phi_2= -\partial_x ( c_u   \phi_2+b_1  (\phi_1^2 - \phi_2^2)  +D_2 \partial_x \phi_2  + B_2 \xi_2)\label{eq:phi2-B}
\end{align}
since the products  $ \phi_1 \phi_3 $ and $ \phi_2 \phi_3 $ become negligibly small due to space separation of the umbilic and single mode. 
After rescaling time as $t \rightarrow \tau=b_1 t$, the reduced  system (\ref{eq:phi1-B}) becomes

\begin{align}
&\partial_\tau \phi_1 = -\partial_x (\frac{c_u}{b_1}   \phi_1 + 2 \phi_1 \phi_2  + D_1' \partial_x \phi_1  + B_1' \xi_1 ) \label{eq:phi1-Brescaled} \\
&\partial_t \phi_2= -\partial_x (\frac{c_u}{b_1}   \phi_2+  (\phi_1^2 - \phi_2^2)  +D_2' \partial_\tau\phi_2  + B_2' \xi_2)\label{eq:phi2-Brescaled}
\end{align}
Comparing the above with the classification of umbilic scenarios  for 2-component systems of Ref.~\cite{2025Spohn}, we find that the 
system lies on the line of fixed points with 
\begin{align}
&X=Y=-1.
\end{align}
Simulations of umbilic mode in the lattice model show good agreement with this prediction, see Fig.~\ref{FigUmbilicLattice}.

\section{\\Effective NLFH equations for an umbilic mode with arbitrary higher degeneracy ($K>2$)}
\label{app:HigherNLFH}
Analogously to the  previous section,  we consider an umbilic mode with arbitrary degeneracy $K$,
appearing on the umbilic line of equal densities $\rho_1=\rho_2 = \ldots =\rho_K =\rho_{K+1} = \rho $. The purpose of this section is to argue that also for higher degeneracy $K>2$ of the umbilic mode, we expect a universal umbilic dynamical structure factor. 
Because of the $K$-degeneracy of the umbilic mode,  the diagonalizing transformation  $R$ has $K-1$ free parameters parameters.  
\begin{align}
&R= (\rho(1-\rho))^{-\frac12} \left(
\begin{array}{c}
 v_1 \\
v_2 \\
 \ldots\\
v_{K+1}
\end{array}
\right), \label{app:Rmulti}
\end{align}
Introducing basic vectors $ e_j$ with elements $(e_j)_k= \de_{j,k}$,  for $j,k=1,2, \ldots K+1$,  
we choose the orthonormal basis $v_j$ in  (\ref{app:Rmulti}) as
\begin{align}
&v_{K+1} =\frac{1}{\sqrt{K+1}} (1,1,\ldots,1) = \frac{1}{\sqrt{K+1}} \sum_{j=1}^{K+1} e_j, \\ 
&v_n =  \frac{1}{\sqrt{n+n^2}}\left( - n e_{n+1} + \sum_{j=1}^{n} e_j \right), \quad n=1,2, \ldots K.
\end{align}
Under the transformation $R$,  the NLFH equations for the umbilic mode $n = 1,2, \ldots K$ take the following form 
\begin{align}
&\partial_t \phi_s = -\partial_x \left(c_u   \phi_s - (s-1)b(K,a,\rho) A_s \phi_s^2 + b(K,a,\rho) A_s \sum_{d=1}^{s-1} \phi_d^2 \right.\no\\
&\left. +2 b(K,a,\rho) \phi_s \sum_{j=s+1}^{K} A_j \phi_j  + D_s \partial_x \phi_s + B_s \xi_s \right) + \mbox {transitory terms},
\label{app:NLFHreduced}
\\
& \quad  s= 1,2, \ldots K, \no
\end{align}
where the transitory terms of type $\phi_s \phi_{K+1}$ 
become negligibly small due to space separation between the umbilic $\phi_s,  \ s\leq K$ and single mode $\phi_{K+1}$ at late times. 
The coefficients $A_s$  in (\ref{app:NLFHreduced}) are constants 
\begin{align}
&A_s = \sqrt{\frac{6}{s(s+1)}},
\end{align}
while $b(K,a,\rho)$ is a function of system parameters,
\begin{align}
&b(K,a,\rho)= -\sqrt{\frac{\rho(1-\rho)}{6}} \left(   1+a+ (K-2)a \rho \right). 
\end{align}
A Galilean transformation $(x,t) \rightarrow (x-c_u t,t )$  removes the terms 
$c_u \partial_x \phi_s$ in (\ref{app:NLFHreduced}).  Then, 
after rescaling time as $t \rightarrow \tau=b(K,a,\rho) t$, the reduced  NLFH system (\ref{app:NLFHreduced})  finally becomes 
\begin{align}
&\partial_\tau \phi_s = -\partial_x \left( - (s-1)A_s \phi_s^2 +  A_s \sum_{d=1}^{s-1} \phi_d^2
+2 \phi_s \sum_{j=s+1}^{K} A_j \phi_j  + \tilde D_s \partial_x \phi_s + \tilde B_s \xi_s \right) 
\label{app:NLFHreduced1}\\
& s= 1,2, \ldots K \no
\end{align}
The  system (\ref{app:NLFHreduced1}) contains no dependence on system parameters $a,\rho$  (dependence on degeneracy $K$
enters through the number of equations),  so  we expect that the two-point correlation function for fixed $K$ is universal.

For $K=2$
the  system (\ref{app:NLFHreduced1}) reduces to Eqs.~(\ref{eq:phi1-Brescaled}) and (\ref{eq:phi2-Brescaled}).  
For $K=3$ and $K=4$  the  NLFH equations take the form 
\begin{align}
&\partial_\tau \phi_1 = -\partial_x \left( 2 \phi_1 \left( \phi_2 + \frac{1}{\sqrt{2}} \phi_3 \right)
+ D_1 \partial_x \phi_1 + B_1 \xi_1 \right) 
\no\\
&\partial_\tau \phi_2 = -\partial_x \left( \phi_1^2- \phi_2^2 +
\sqrt{2}\, \phi_2  \phi_3  
+ D_2 \partial_x \phi_2 + B_2 \xi_2 \right) \label{app:NLFHreducedK4} \\
&\partial_\tau \phi_3 = -\partial_x  \left( \frac{1}{\sqrt{2}} \left(\phi_1^2+ \phi_2^2 -2  \phi_3^2 \right)
+ D_3 \partial_x \phi_3 + B_3 \xi_3 \right) \no 
\end{align}
and 
\begin{align}
&\partial_\tau \phi_1 = -\partial_x \left( 2 \phi_1 \left( \phi_2 + \frac{1}{\sqrt{2}} \phi_3 + \sqrt{\frac{3}{10}} \phi_4 \right)
+ D_1 \partial_x \phi_1 + B_1 \xi_1 \right) \no
\\
&\partial_\tau \phi_2 = -\partial_x \left( \phi_1^2- \phi_2^2 +
2 \phi_2 \left(  \frac{1}{\sqrt{2}} \phi_3 + \sqrt{\frac{3}{10}} \phi_4 \right)
+ D_2 \partial_x \phi_2+ B_2 \xi_2 \right) \label{app:NLFHreducedK4} \\
&\partial_\tau \phi_3 = -\partial_x  \left( \frac{1}{\sqrt{2}} \left(\phi_1^2+ \phi_2^2 -2  \phi_3^2 \right)
+2 \phi_3 \left( \sqrt{\frac{3}{10}} \phi_4 \right)
+ D_3 \partial_x \phi_3 + B_3 \xi_3 \right) \no \\
&\partial_\tau \phi_4 = -\partial_x  \left(\sqrt{\frac{3}{10}} \left(\phi_1^2+ \phi_2^2+ \phi_3^2 -3  \phi_4^2 \right)
+ D_4 \partial_x \phi_4 + B_4 \xi_4 \right).\no 
\end{align}

\end{document}